\documentclass[12pt,a4paper]{article}
\usepackage{graphics}
\usepackage{epsf,cite,a4wide}
\newcommand{\beq}{\begin{equation}}
\newcommand{\eeq}{\end{equation}}
\newcommand{\bea}{\begin{eqnarray}}
\newcommand{\eea}{\end{eqnarray}}
\renewcommand{\d}{\delta}
\renewcommand{\l}{\lambda}

\renewcommand{\b}{\beta}

\renewcommand{\ni}{\noindent}

\newcommand{\m}{\mu}

\newcommand{\s}{\sigma}

\newcommand{\W}{{\cal W}}
\renewcommand{\th}{\theta}

\newcommand{\oh}{\frac{1}{2}}
\newcommand{\oq}{\frac{1}{4}}

\newcommand{\dg}{\dagger}
\newcommand{\non}{\nonumber}

\newcommand{\rf}[1]{(\ref{#1})}
\newcommand{\ra}{\rightarrow}

\begin{document}

\hfill January 1998

\begin{center}

\vspace{18pt}

  {\Large \bf Detection of Center Vortices in the \\
      Lattice Yang-Mills Vacuum }

\end{center}

\vspace{18pt}

\begin{center}
{\sl L. Del Debbio${}^a$, M. Faber${}^b$, J. Giedt${}^c$, J. Greensite${}^d$,
and {\v S}. Olejn\'{\i}k${}^e$}

\end{center}

\bigskip

\begin{tabbing}

{}~~~~~~~~~~~~~~~~~~~~\= blah  \kill
\> ${}^a$ Dept. of Physics and Astronomy, Univ.\ of Southampton, \\
\> ~~Southampton SO17 1BJ, UK.  E-mail: {\tt ldd@hep1.phys.soton.ac.uk} \\
\\
\> ${}^b$ Inst. f\"ur Kernphysik, Technische Universit\"at Wien, \\
\> ~~A-1040 Vienna, Austria.  E-mail: {\tt faber@kph.tuwien.ac.at} \\
\\
\> ${}^c$ Physics Department, University of California, Berkeley,\\
\> ~~CA 94720 USA.  E-mail: {\tt giedt@socrates.berkeley.edu} \\
\\
\> ${}^d$ The Niels Bohr Institute, DK-2100 Copenhagen \O, \\
\> ~~Denmark.  E-mail: {\tt greensite@nbivms.nbi.dk} \\
\\
\> ${}^e$ Institute of Physics, Slovak Academy of Sciences, \\
\> ~~SK-842 28 Bratislava, Slovakia.  E-mail: {\tt fyziolej@savba.sk}

\end{tabbing}

\vspace{18pt}

\begin{center}

{\bf Abstract}

\end{center}

\bigskip

   We discuss the implementation of the ``direct'' maximal center
gauge (a gauge which maximizes the lattice average of the squared-modulus
of the trace of link variables), and its use in identifying $Z_2$ center 
vortices in Yang-Mills vacuum configurations generated by lattice Monte 
Carlo.  We find that center vortices identified in the vacuum state account 
for the full asymptotic string tension.  Scaling of vortex densities
with lattice coupling, change in vortex size with cooling, and 
sensitivity to Gribov copies is discussed.  Preliminary evidence is presented,
on small lattices, for center dominance in $SU(3)$ lattice gauge theory. 

\vfill

\newpage

\section{Introduction}

     In a number of recent articles and conference proceedings
\cite{PRD97,Zako,lat97} we have presented numerical evidence in favor
of the Center Vortex theory of confinement, which was put forward in
the late 1970's \cite{tHooft,Mack,Cop,Corn,Feyn}. Our most important
tool is the use of the maximal center gauge, which, combined with
``center projection,'' allows us to identify the locations of center
vortices in thermalized lattice gauge-field configurations. It is found
that these vortices, by themselves, account for the entire asymptotic
string tension (``center dominance'').  We have also found evidence
\cite{Zako} that the monopoles of the maximum abelian gauge lie along
the center vortices in a monopole-antimonopole chain, and that their
non-abelian field strength, above the vacuum average, is almost
entirely oriented in the vortex direction.  This opens the way to
explain abelian dominance, and monopole condensation, in terms of more
fundamental underlying vortex configurations.  Finally, in ref.
\cite{castex}, we have argued that the Casimir scaling of higher
representation string tensions, formerly a very strong argument
\emph{against} the center vortex theory, can in fact be understood in
terms of center vortices.  Independent arguments in favor of the
center vortex theory have been presented by Kov\'{a}cs 
and Tomboulis \cite{TK}, who follow a rather different approach
but reach similar conclusions.
           
    In this article we will explain, in section 2, the actual implementation 
of the ``direct'' maximal center gauge, which underlies much of our work,
and review the evidence, in the direct gauge, that center vortices are 
responsible for quark confinement.\footnote{Some of this evidence was 
obtained previously in a slightly different, ``indirect'', version of 
maximal center gauge \cite{PRD97,Zako}.}  We go on to study 
(section 3) how vortex configurations are affected by cooling the lattice, 
and then take up issues related to Gribov copies 
(section 4).  Some preliminary evidence for center dominance in 
$SU(3)$ lattice gauge theory is presented in section 5, followed by
a summary of our results in section 6.    
   
\section{Center Vortices and Confinement}

   Our procedure for locating center vortices in thermalized lattice
configurations was inspired by earlier work
on abelian projection in maximal abelian gauge \cite{Suz}.  The idea
is to fix a gauge (the maximal center gauge) which, in the case of
$SU(2)$ gauge theory, reduces the full $SU(2)$ gauge symmetry to the center 
subgroup $Z_2$.  ``Center projection'' is a mapping of the full gauge
field configuration $U_\m(x)$ onto a configuration $Z_\m(x)=\pm 1$,
transforming as a $Z_2$ gauge field under the residual $Z_2$ symmetry.
The excitations of a $Z_2$ gauge field are (thin) center vortices, and these
are used to locate thick center vortices in the full, unprojected gauge-field
configuration $U_\m(x)$, as explained below.

    In fact we have introduced two versions of the maximal center gauge;
an ``indirect'' version, in refs. \cite{lat96,PRD97}, and a ``direct''
maximal center gauge in ref. \cite{Zako}.  The indirect gauge is a
further gauge-fixing within maximal abelian gauge, which reduces the residual
$U(1)$ gauge symmetry to $Z_2$.  One begins by fixing to maximal abelian 
gauge, defined as the gauge which maximizes
\beq
       Q = \sum_{x,\m} \mbox{Tr}[U_\m(x)\s_3 U_\m^\dg(x) \s_3]
\eeq
leaving a residual $U(1)$ symmetry.  This gauge makes the link variables
as diagonal as possible.   Abelian link variables are
extracted from the diagonal elements of the full link variables, rescaled
to restore unitarity:
\bea
      A &=& {\mbox{diag}[U_{11},U_{22}] \over 
             \sqrt{U_{11}U^*_{11} + U_{22}U^*_{22}} }
\non \\
        &=& \left[ \begin{array}{cc}
                      e^{i\th} &    \cr
                               & e^{-i\th} \end{array} \right]
\eea
These variables transform under the residual symmetry like $U(1)$ gauge field
link variables.  The indirect maximal center gauge uses this residual symmetry
to bring the $A$-link variables as close as possible to $SU(2)$
center elements $\pm I$, by maximizing the lattice average of
$\cos^2(\th)$.  Center projection is then achieved by identifying
\beq
        Z_\m(x) = \mbox{sign}[\cos(\th_\m(x))]
\eeq
The gauge is ``indirect'' in the sense that the center is maximized
in the abelian link variable $A$, rather than directly in the full link
variable $U$.  String tensions can be extracted from the center-projected
configurations and, although agreement with the asymptotic string
tension of the full configurations is not too bad, significantly better
results are obtained in the direct maximal center gauge, as will be seen
below.

   The direct maximal center gauge, in $SU(2)$ gauge theory, is defined as 
the gauge which brings the full link variables $U$ as close as possible to 
the center elements $\pm I$, by maximizing the quantity
\beq
       R = \sum_{x,\m} \mbox{Tr}[U_\m(x)]^2
\label{R}
\eeq
with center projection defined by
\beq
       Z_\m(x) = \mbox{sign}[\mbox{Tr}U_\m(x)]
\label{cp}
\eeq
Again, the projected $Z_\m(x)$ field transforms as a gauge field under
the residual $Z_2$ symmetry.  Before going on to discuss numerical results 
obtained in this gauge, we must first discuss how to implement it.

\subsection{Fixing to direct maximal center gauge}

   The gauge-fixing is accomplished by over-relaxation \cite{MO}.  
Beginning with a thermalized but non-gauge-fixed lattice, we sweep 
through the lattice site by site.  At each site $x$, one needs to find the 
gauge transformation $g$ which maximizes the local quantity
\beq
        R_x = \oq \left\{ \sum_{\m} \mbox{Tr}[g(x)U_\m(x)]^2
          + \sum_{\m} \mbox{Tr}[U_\m(x-\hat{\m})g^\dg(x)]^2 \right\}
\eeq
Denote
\bea
        g(x) &=& g_4 I - i \vec{g} \cdot \vec{\s}
\non \\
        U_\m(x) &=& d_4(\m) I + i \vec{d}(\m) \cdot \vec{\s}
\non \\
 U_\m(x-\hat{\m}) &=& d_4(\m+4) I - i \vec{d}(\m+4) \cdot \vec{\s}
\eea
where the choice of signs in front of the various terms proportional
to $i$ is made for convenience.  Then
\beq
        R_x = \oh \sum_{l=1}^8 \left( \sum_{k=1}^4 g_k d_k(l) \right)^2
\eeq
We have to maximize this quantity subject to the constraint that $g$
is unitary.  To this end, introduce a Lagrange multiplier
\beq
      \tilde{R} = R_x + \oh \l \left( 1 - \sum_{k=1}^4 g_k^2 \right)
\eeq
Then the conditions for a maximum satisfying the constraint, obtained
by differentiating $\tilde{R}$, are
\bea
     \sum_{j=1}^4 \sum_{l=1}^8 d_i(l) d_j(l) g_j &=& \l g_i
\non \\
            \sum_{k=1}^4 g_k^2 = 1
\eea
This can be written as an eigenvalue equation
\beq
            D \vec{G} = \lambda \vec{G}
\eeq
where
\beq
            D_{ij} = \sum_{l=1}^8 d_i(l) d_j(l)
\eeq
and the unitarity constraint is the normalization condition
\beq
         \vec{G} \cdot \vec{G} = 1
\eeq
At this point, the problem of finding $g$ boils down to finding the
eigenvectors of a $4\times 4$ real symmetric matrix, which is 
achieved by standard methods.  There are four eigenvectors corresponding
to four stationary points.  The eigenvector with the largest eigenvalue
corresponds to the gauge transformation $g$ maximizing $R_x$ at site $x$.

   The next step is to apply the over-relaxation algorithm.  We transform
the links touching site $x$ not by $g(x)$ but by $g^\omega(x)$, where
\bea
       g &=& g_4 I - i \vec{g}\cdot \vec{\s}
\non \\
         &=& \cos(\phi) I - i \vec{n}\cdot \vec{\s} \sin(\phi)
\non \\
      g^\omega &=& \cos(\omega \phi) I - i \vec{n}\cdot \vec{\s} 
                                              \sin(\omega \phi)
\eea
and we use $\omega=1.7$ \cite{MO}.  This procedure is
applied at each site of the lattice, sweeping through the lattice several 
hundred times.  The algorithm stops when a convergence criterion is
satisfied.  Our criterion is that the lattice average of $(\oh \mbox{Tr}U)^2$
changes by less than $0.00015$ after 50 gauge-fixing sweeps.

   Finally, it must be noted that maximal center gauge, like
the Coulomb, Landau, and maximal abelian gauges, is afflicted with
Gribov copies.  To alleviate the problem,
we make several gauge copies of the original 
configuration by applying random gauge transformations, and then gauge fix 
each copy.  The gauge-fixed copy with the largest value of $R$ is chosen for 
data-taking.  In practice, three gauge copies seems sufficient, in the sense 
that we don't improve the average value of $R$ very much (or change the final 
results) by making more copies.  

  All of the data reported below was obtained in the direct maximal
center gauge.  A portion of those results, displayed in Figures 
\ref{nvtex}-\ref{nchi} below, are similar to results obtained
previously in the indirect version of maximal center gauge \cite{PRD97}.

\subsection{Projection vortices locate center vortices}

   Having fixed to the direct maximal center gauge by this procedure,
we obtain the corresponding center-projected configuration 
(a $Z_2$ lattice gauge field), from \rf{cp}.  The excitations of any
$Z_2$ gauge field are line-like (D=3) or surface-like (D=4) vortices
on the dual lattice.  We refer to these excitations,
in the center-projected configurations, as ``projection vortices'' or
just ``P-vortices.''  The question is whether P-vortices in the projected
configuration are in any way related to the existence of center vortices
in the full, unprojected lattice configuration.

   In order to study this question, we introduce the concept of
vortex-limited Wilson loops $W_n(x)$.   We say that a plaquette 
is pierced by a P-vortex if, upon going to maximal center gauge and 
center-projecting, the projected plaquette has the value $-1$.
Likewise, a given lattice surface is pierced by $n$ P-vortices if 
$n$ plaquettes of the surface are pierced by P-vortices.  As a Monte
Carlo simulation proceeds, the number of P-vortices piercing any
given surface will vary.  
Define $W_n(C)$ to be the Wilson loop evaluated on a sub-ensemble
of configurations, selected such that precisely $n$ P-vortices, in
the corresponding center-projected configurations, pierce the minimal area
of the loop.  It should be emphasized here that the center projection
is used only to select the data set.  The Wilson loops themselves are
evaluated using the full, \emph{unprojected} link variables.  In practice,
to compute $W_n(C)$, the procedure is to generate thermalized lattice 
configurations by the usual Monte Carlo algorithm, and fix to maximal
center gauge as described above.  For each independent configuration
one then examines each rectangular loop 
on the lattice of a given size; those with $n$ P-vortices piercing the
loop are evaluated, the others are skipped.
         
   The test for whether (thin) P-vortices in the projected configuration 
correspond to (thick) center vortices in the full, 
unprojected $SU(2)$ gauge-field configuration is whether the behavior
\beq
       {W_n(C) \over W_0(C)} \ra (-1)^n
\label{ratio1}
\eeq
is found in the limit of large loop area.  
The reasoning behind this test has 
been given elsewhere \cite{PRD97,Zako}, but for completeness we repeat the
argument here.

    Vortices are created by discontinuous gauge transformations.  Suppose 
loop $C$, pa\-ra\-met\-rized by $x^\mu(\tau),~\tau \in [0,1]$, 
encircles $n$ vortices.  
At the point of discontinuity
\begin{equation}
       g(x(0)) = (-1)^n g(x(1))
\label{discont}
\end{equation}
The corresponding vector potential, in the neighborhood of loop $C$ can
be decomposed as
\begin{equation}
       A^{(n)}_\mu(x) = g^{-1}\delta A^{(n)}_\mu(x) g + i g^{-1} 
\partial_\mu g
\label{back}
\end{equation}
with the inhomogenous term dropped at the point of discontinuity. Then
\begin{eqnarray}
       W_n(C) &=& <\mbox{Tr}\exp[i\oint dx^\mu A_\mu^{(n)}]>
\nonumber \\ 
          &=& (-1)^n <\mbox{Tr}\exp[i\oint dx^\mu \delta A_\mu^{(n)}]> 
\end{eqnarray}
In the region of the loop $C$, the vortex background looks locally like
a gauge transformation.  If all other fluctuations $\delta A^{(n)}_\mu$ are
basically short-range, then they should be oblivious, in the neighborhood
of the loop $C$, to the presence or absence of vortices in the middle of
the loop.  In that case, if we have correctly identified the vortex
contribution, then
\begin{equation}
<\mbox{Tr}\exp[i\oint dx^\mu \delta A_\mu^{(n)}]> ~ \approx ~
      <\mbox{Tr}\exp[i\oint dx^\mu \delta A_\mu^{(0)}]> 
\end{equation}
for sufficiently large loops, and eq. \rf{ratio1} follows
immediately.  The validity of eq. \rf{ratio1} then constitutes a test of
whether P-vortices, which we are using to select the subscript $n$ of 
$W_n(C)$, actually locate center vortices in the unprojected configurations. 

\begin{figure}
\centerline{\scalebox{.7}{\includegraphics{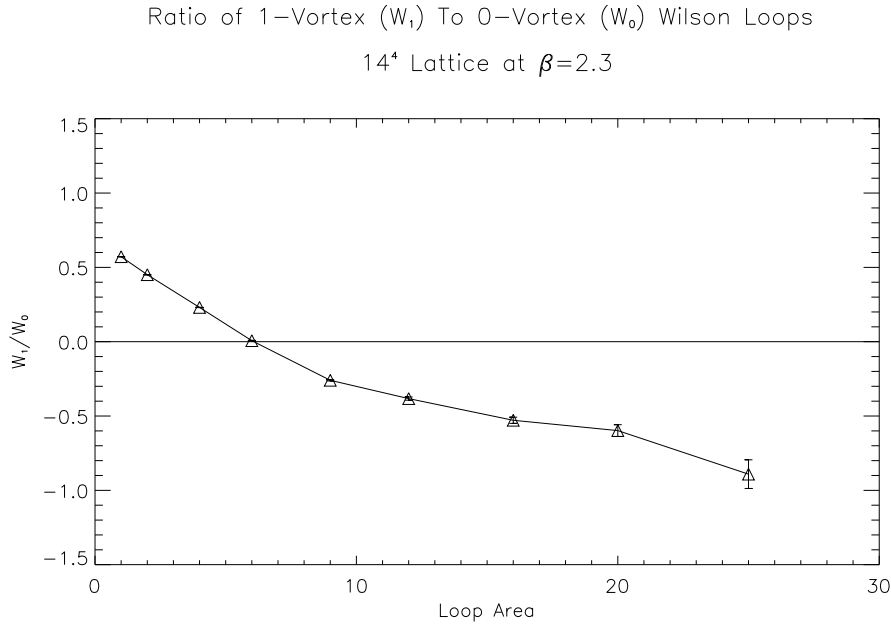}}}
\caption{Ratio of the 1-Vortex to the 0-Vortex Wilson loops, $W_1(C)/W_0(C)$, 
vs.\ loop area at $\b=2.3$.}
\label{nvtex}
\end{figure}

\begin{figure}
\centerline{\scalebox{.7}{\includegraphics{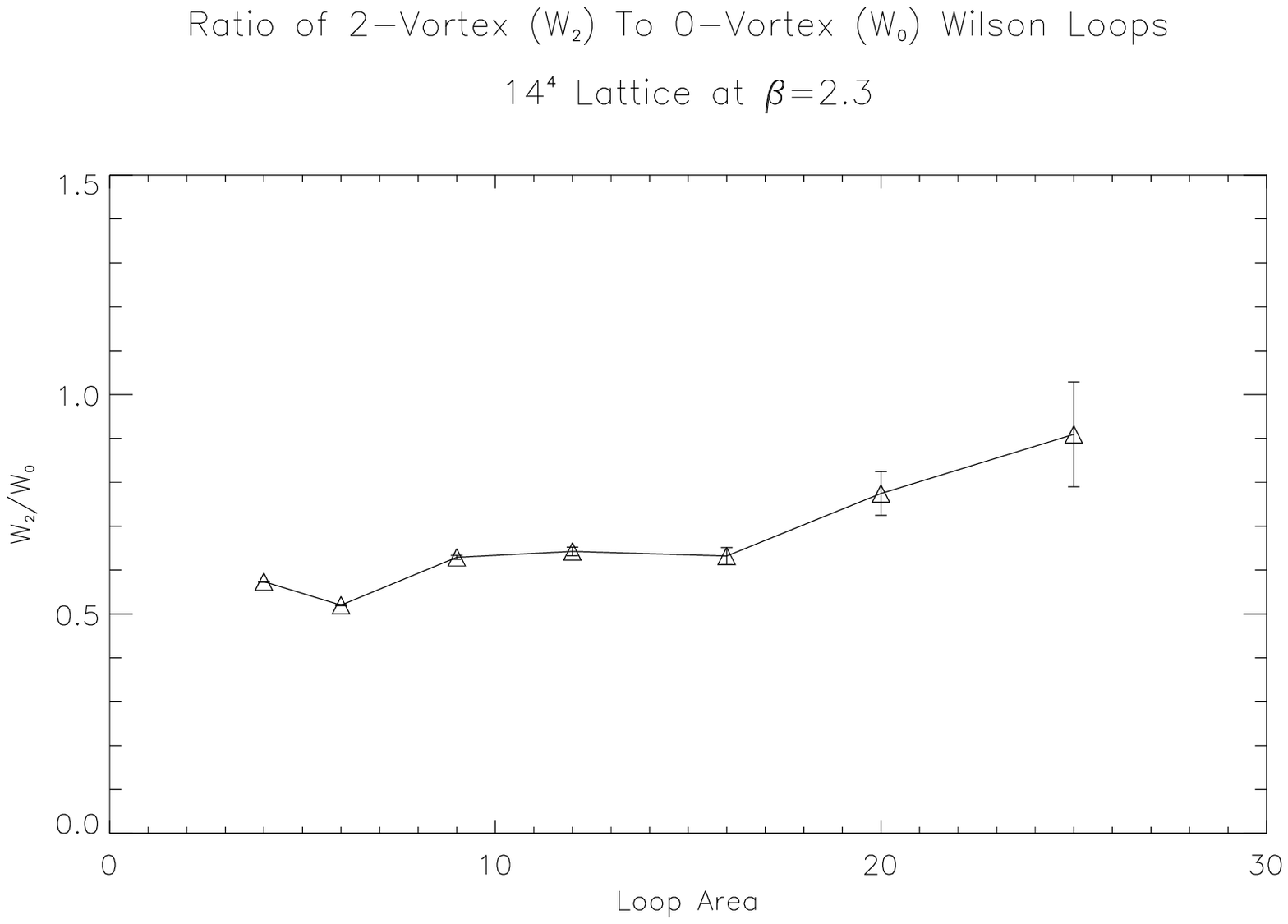}}}
\caption{Ratio of the 2-Vortex to the 0-Vortex Wilson loops, $W_2(C)/W_0(C)$, 
vs.\ loop area at $\b=2.3$.}
\label{nvtex2}
\end{figure}

   Figure \ref{nvtex} shows the ratio $W_1/W_0$, 
with the single P-vortex associated with $W_1$ located at (or touching)
the center of the loop.  Likewise, Fig. \ref{nvtex2} shows
the ratio $W_2/W_0$, with the two P-vortices for $W_2$ located near the 
center of the loop.  Both figures were obtained from a simulation on a
$14^4$ lattice at $\b=2.3$ (400 configurations separated by 100 sweeps), and 
both appear to be quite consistent with the limiting
behavior \rf{ratio1}.\protect\footnote{Qualitatively similar results were found
in the indirect maximal center gauge \cite{PRD97}.}     

\begin{figure}
\centerline{\scalebox{.7}{\includegraphics{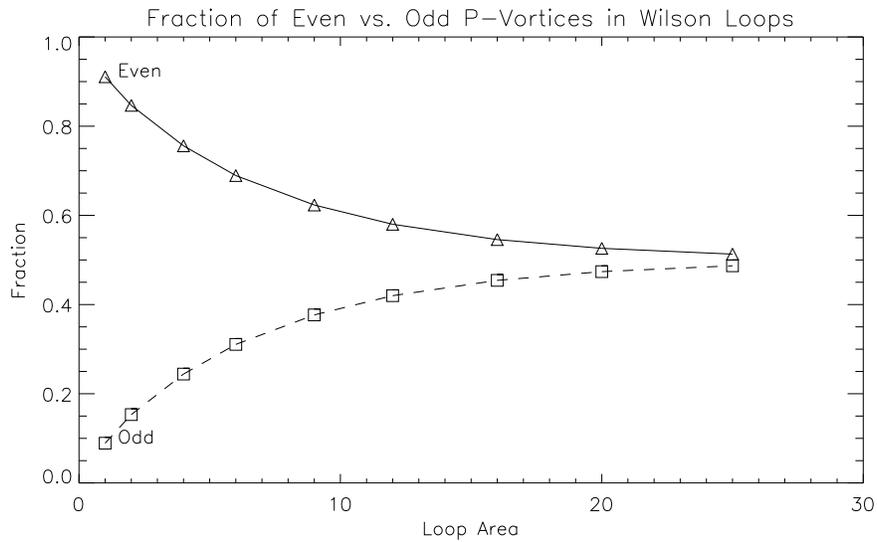}}}
\caption{Fraction of link configurations containing even (or zero)/odd 
numbers of P-vortices, at $\b=2.3$, piercing loops of various areas.}
\label{nfrac}
\end{figure}

\begin{figure}
\centerline{\scalebox{.7}{\includegraphics{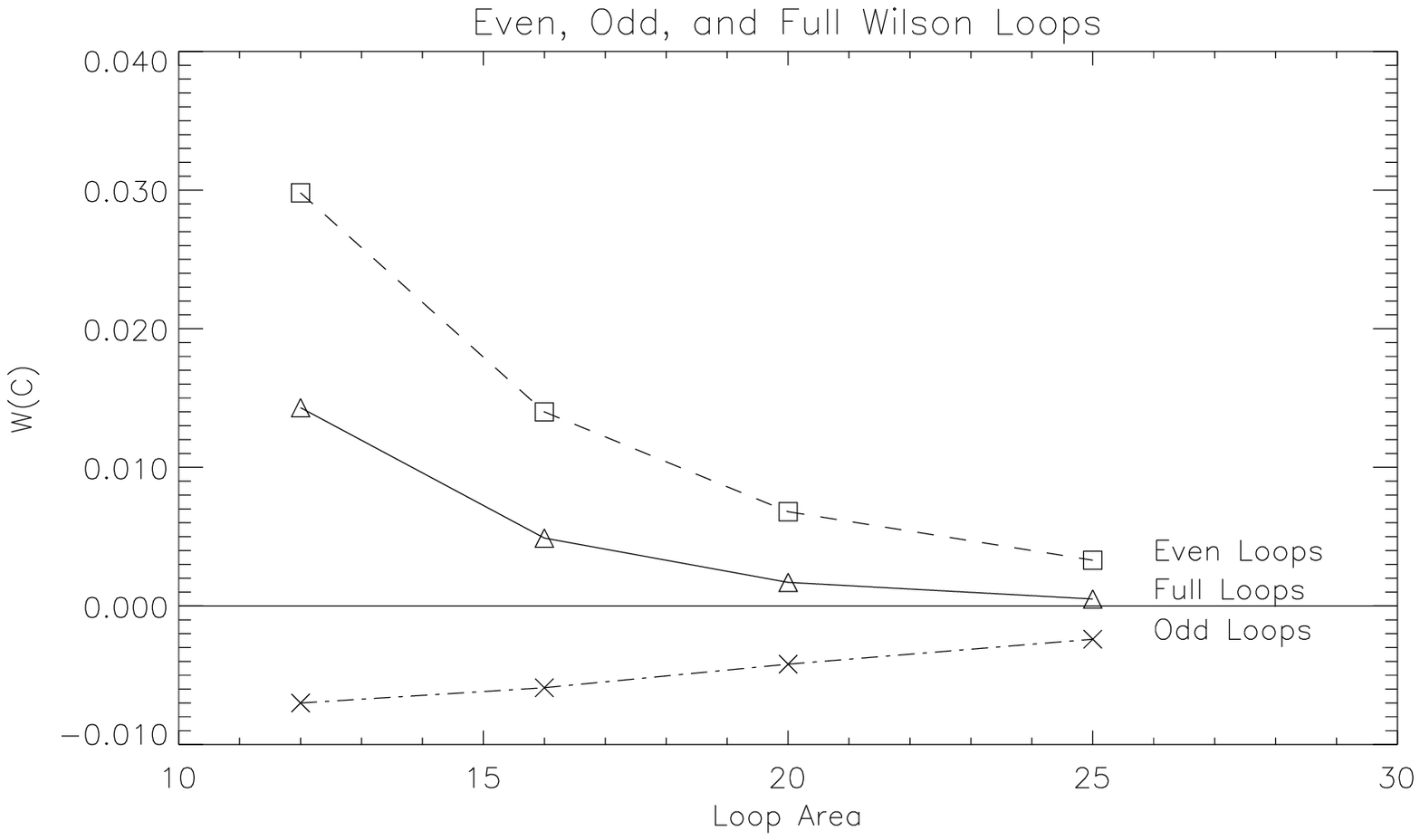}}}
\caption{Wilson loops $W_{evn}(C),~ W_{odd}(C)$, and $W(C)$ at larger
loop areas, extracted from configurations with even (or zero) numbers
of P-vortices, odd numbers of P-vortices, any any number of P-vortices,
respectively, piercing the minimal loop area, again at $\b=2.3$.}
\label{nvtexeo}
\end{figure}

   For very large loops, the fraction of configurations in which no
vortex pierces the loop (the subensemble used to compute $W_0$) becomes
very small. So as a further check, using \emph{all} the configurations, 
we define $W_{evn}(C)$ to be the Wilson loop evaluated in sub-ensemble in 
which only even (including zero) numbers of P-vortices pierce the minimal area,
while $W_{odd}(C)$ is the corresponding quantity for odd numbers of
P-vortices. For a very large loop, the fraction of configurations used
to evaluate $W_{evn}(C)$, denoted $P_{evn}(C)$, and the fraction 
$P_{odd}(C)$ used to evaluate $W_{odd}(C)$, should each approach $50\%$
of the total configurations. This is in fact the case, as seen in
Fig.\ \ref{nfrac}. If P-vortices in the projected lattice are associated 
with center vortices in the unprojected lattice, then we would expect,
by the same argument leading to eq. \rf{ratio1}, that
\newpage
\beq
       {W_{odd}(C) \ra - W_{evn}(C)} 
\label{ratio2}
\eeq
in the limit of large loop area. That also appears to be the case, as seen 
in Fig.\ \ref{nvtexeo}.  

   The conclusion is that P-vortices in center-projected lattice
configurations obtained in direct maximal center gauge serve to 
locate thick center vortices in the full, unprojected, lattice gauge 
field configuration.  It is well to bear in mind, however, that we have no
real understanding of \emph{why} this technique finds center vortices; our 
confidence is based entirely on the numerical results shown in this and the 
following sections.

\subsection{No vortices means no area law}

\begin{figure}
\centerline{\scalebox{.7}{\includegraphics{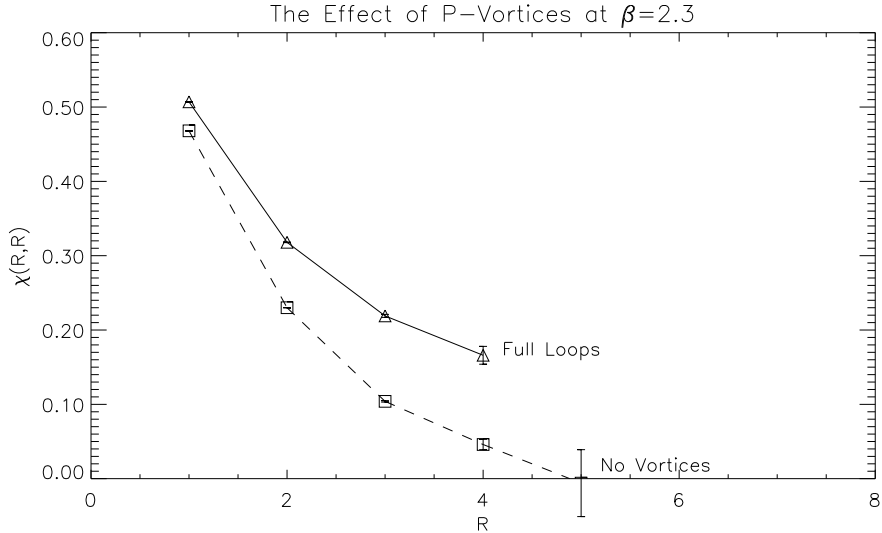}}}
\caption{Creutz ratios $\chi_0(R,R)$ extracted from loops with no
P-vortices, as compared to the usual Creutz ratios $\chi(R,R)$ at
$\b=2.3$.}
\label{nchi0}
\end{figure}

   The fact that center vortices can be identified in the gauge-field 
vacuum does not necessarily imply that vortices 
are important for the confinement
mechanism.  There is, however, a simple test of their relevance.  Let us define
$\chi_n(I,J)$ as the Creutz ratio extracted from the vortex-limited
loops $W_n(C)$.  If the presence or absence of center vortices crossing
the minimal spanning surface of a loop is unrelated to the area-law falloff,
then we would naturally expect, at least for large loops, that     
\beq
       \chi_0(I,J) \approx \chi(I,J)
\eeq
where $\chi(I,J)$ is the usual Creutz ratio with no restriction on numbers
of vortices.  In fact, the above equation is entirely wrong, as seen
in Fig. \ref{nchi0}.  When Wilson loops are evaluated in subensembles in which
no vortices cross the minimal area of the loop, the string tension 
vanishes.  
      
\begin{figure}
\centerline{\scalebox{.7}{\includegraphics{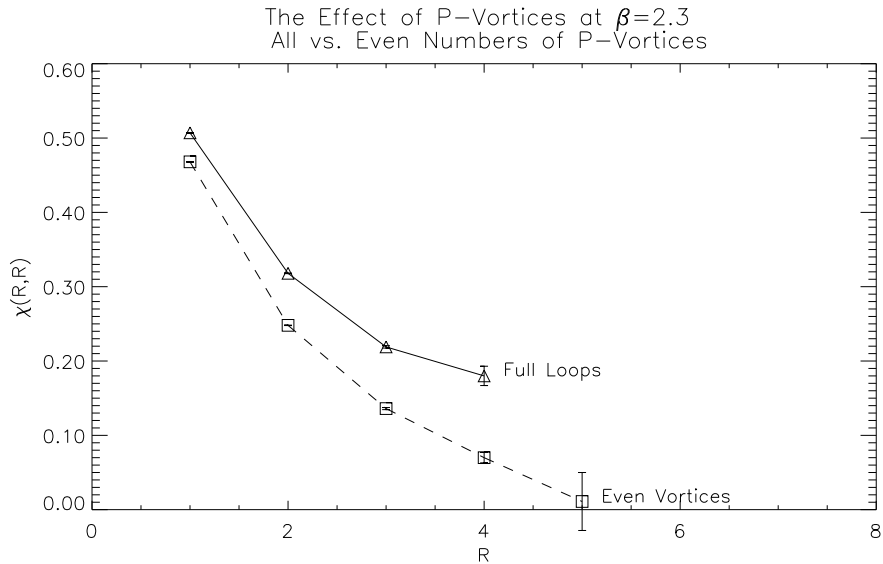}}}
\caption{Creutz ratios $\chi_{evn}(R,R)$ extracted from loops 
$W_{evn}(C)$, taken from configurations with even (or zero) numbers
of P-vortices piercing the loop.  The standard Creutz ratios at
this coupling ($\b=2.3$) are also shown.}
\label{nchieo}
\end{figure}

   As a further test, we may consider loops pierced by even (or zero)
numbers of vortices.  According to the center vortex theory, the asymptotic 
string tension is entirely due to fluctuations in the number of center vortices
piercing the surface of the loop.  The asymptotic effect of creating
a vortex piercing (once or an odd number of times) the loop surface
is to multiply the loop by a center element, i.e. in $SU(2)$
\beq
       \mbox{Tr}[UU...U] \ra (-1) \times  \mbox{Tr}[UU...U]
\eeq
and therefore the area-law falloff is due, asymptotically, to a delicate 
cancellation between configurations with even (or zero) numbers of vortices 
piercing the loop (which gives a positive average contribution), and
configurations with odd numbers of vortices piercing the loop (which gives
a negative average contribution).  In fact, we have already seen some evidence
of this cancellation in Fig.\ \ref{nvtexeo}, where the value of the full
loop $W(C)$ is much smaller than the magnitudes of either the
even or the odd components $W_{evn}(C),~W_{odd}(C)$.  More quantitatively,
if we evaluate Creutz ratios $\chi_{evn}(I,J)$ evaluated from the 
even-vortex ($W_{evn}$) contribution alone, the vortex theory predicts that
\beq
         \chi_{evn}(I,J) \ra 0
\eeq
in the limit of large loop area.  Once again, from Fig.\ \ref{nchieo},
this appears to be exactly what happens.
 
\subsection{Center vortices generate the full asymptotic \\ string tension}

    From the previous results, we deduce that
the confining properties of lattice gauge-field configurations
are strongly correlated with distribution of center vortices.  The final
check is whether these vortices account for the \emph{entire}
asymptotic string tension, as predicted by the center vortex theory.
We have already seen that the asymptotic effect of $n$ vortices piercing
the middle of a large loop is to contribute a factor $(-1)^n$ to the
loop value.  In that case, the expectation value of a large Wilson loop can
be factored into two components: (i) a factor $W_{vor}(C)$ due to the effect
of vortices crossing the minimal area, far from the perimeter of the loop;
and (ii) a factor $W_{per}(C)$ due to short-range fluctuations 
(denoted $\d A^{(n)}$ in eq. \rf{back}) around
the vortex background, near the loop perimeter.  Asymptotically, for
large-area loops, the vortex theory predicts that
\bea
       W(C) &\ra& W_{vor}(C) W_{per}(C)
\non \\
       W_{vor}(C) &=& <(-1)^n>  ~~~ n = \mbox{no. of vortices piercing C}
\eea
Since $W_{per}(C)$ should behave asymptotically like $W_{evn}(C)$ or
$|W_{odd}(C)|$, it does not have an area-law falloff, and the entire string 
tension must be due to $W_{vor}(C)$.  But if, as we have seen in the 
previous section, P-vortices locate center vortices, then
\beq
      W_{vor}(C) = <ZZZ...Z> ~~~ \mbox{(center-projected loop value)}
\eeq
where the product of links $ZZ...Z$ on the projected lattice is taken around 
loop $C$. Therefore,

\begin{itemize}
\item{\emph{if} P-vortices locate center vortices, and}
\item{\emph{if} the center vortex theory is correct;}
\item{\emph{then} the string tension of center projected loops should
exactly match the asymptotic string tension of the full theory.}
\end{itemize}

\begin{figure}[h]
\centerline{\scalebox{.7}{\rotatebox{90}{\includegraphics{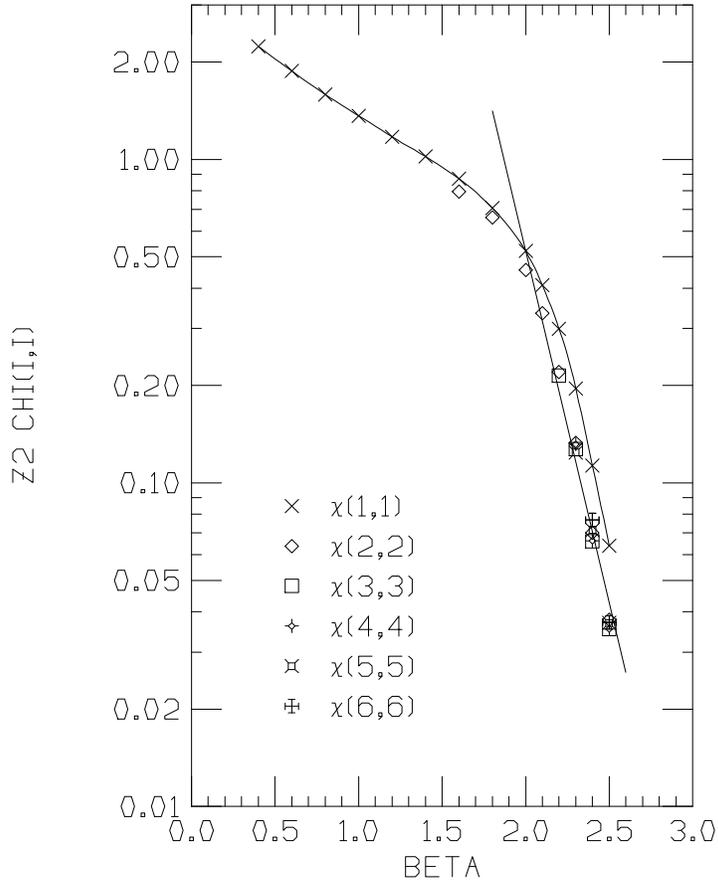}}}}
\caption{Creutz ratios from center-projected lattice configurations,
in the direct maximal center gauge.}
\label{nchi}
\end{figure}

   Figure \ref{nchi} is a Creutz ratio plot, extracted from center-projected
Wilson loops (i.e. from loops of the $Z_\m(x)$ link variables) in
direct maximal center gauge.  The straight solid line is the usual
two-loop expression ($\s=$ string tension, $a=$ lattice spacing)
\beq
      \s a^2 = {\s \over \Lambda^2} \left( {6\pi^2 \over 11} \b 
      \right)^{102/121} \exp\left[- {6\pi^2 \over 11} \b \right]
\eeq
with $\s / \sqrt{\Lambda} = 58$.  There are two aspects of this plot
which are worth noting in particular.  First, unlike a standard plot
in the unprojected theory, the Creutz ratios almost fall on top
of one another, starting at $R=2$.  This is not so surprising, from the
point of view of the vortex theory.  The short range gluonic fluctuations 
which give rise to the Coulombic potential have been eliminated (these
would contribute to $W_{per}(C)$); only the fluctuations in vortex number,
which give rise to a linear potential, remain.  Second, even $\chi(1,1)$,
which is just the logarithm of the center-projected plaquette, appears
to be scaling.  This fact, as we will see, is related to
the scaling of the vortex density. 

\begin{figure}[h]
\centerline{\scalebox{.6}[.9]{\includegraphics{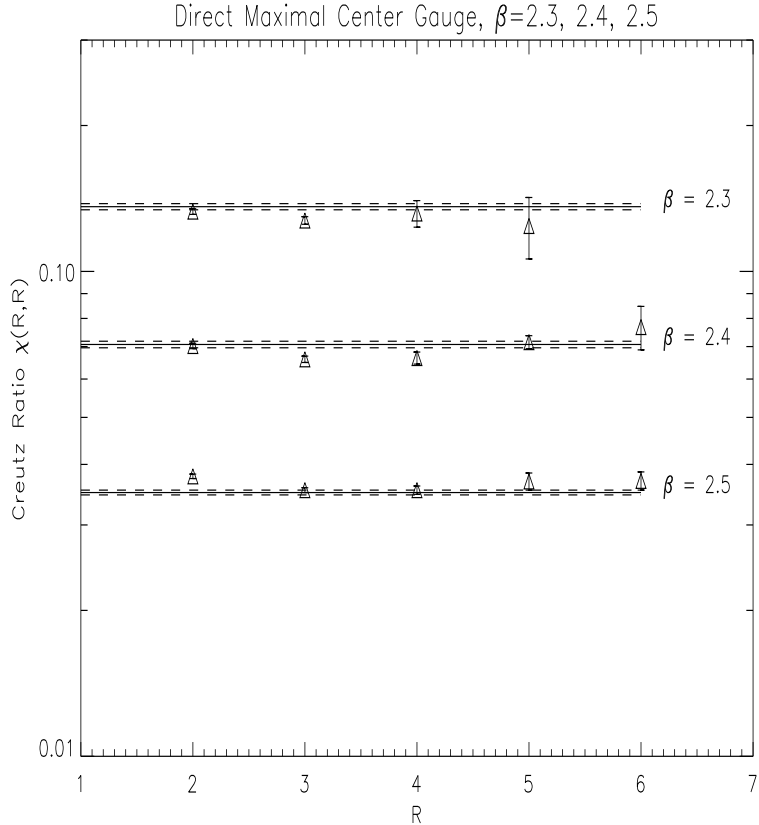}}}
\caption{Center-projection Creutz ratios $\chi(R,R)$ vs.\ R at 
$\b=2.3,~2.4,~2.5$.  Triangles are our data points.  The solid line shows 
the value (at each $\b$) of the asymptotic string tension of the unprojected 
configurations, and the dashed lines the associated error bars, quoted in 
ref.\ \cite{Bali}.}
\label{ynot}
\end{figure}

   Scaling of the center-projected string tension is not sufficient
for our purposes; what is necessary is that the actual \emph{value}
of the string tension, at every $\beta$, agrees with the value for the
asymptotic string tension of the unprojected configurations.
This is also what we find.  
Figure \ref{ynot} shows our data (triangles) in the scaling region, 
for Creutz ratios $\chi(R,R)$ of center-projected Wilson loops, as
compared to the value for the full theory of the asymptotic
string tension.\footnote{This data was taken, for $\b=2.3$ and $\b=2.4$, 
on $16^4$ lattices with 30 configurations separated by 100 sweeps.  
A $22^4$ lattice and 20 configurations separated by 100 sweeps was used 
for $\b=2.5$.}  The values for the full theory (solid lines), with associated
error bars (dashed lines), are taken from Bali et al.\ \cite{Bali}.  This 
agreement of the center-projected and full asymptotic string tension 
persists into the strong coupling regime (see Fig.\ \ref{SU2strong} below).  

\subsection{Scaling of the center vortex density}

   Finally, we consider the density of vortices.  In the vortex theory
of confinement, vortices must be ``condensed'' in the sense that the average
extension of a vortex is on the order of the lattice size itself.  Its easy
to see why.  Suppose the opposite were true, i.e. that there were some
upper limit to vortex extension, and that almost all vortices, in a very
large lattice, would fit inside a hypercube of side length $L$.  Now consider
$R\times T$ Wilson loops with $R,T >> L$.  Then only vortices within a
distance $L$ of the loop perimeter could be linked to the loop, and this
would lead asymptotically to a perimeter-law, rather than area-law, falloff.   

   If vortices are physical objects, their density should scale with $\b$ 
in some appropriate way.  P-vortices are located somewhere near the middle
of ``thick'' center vortices in the unprojected lattice, and these P-vortices 
have the topology of surfaces in D=4 dimensions.  If center vortices scale
correctly, P-vortices should also scale.   The proper asymptotic scaling 
of P-vortex densities in the \emph{indirect} maximal center gauge was reported 
recently by Langfeld et al.\ \cite{LR}.  We can also observe this 
scaling in the direct center gauge in a rather simple way:  We first define 
$p$ to be the fraction, and $N_{vor}$ to be the total number, of center 
projected 
plaquettes with value $-1$.  $N_{vor}$ is also the total area of all 
P-vortices on the dual lattice, and we denote by $N_T$ the total number of 
all plaquettes on the lattice.  Then
\bea
           p &=& {N_{vor} \over N_T} = {N_{vor} a^2 \over N_T a^4} a^2
\non \\
             &=& {\mbox{Total Vortex Area} \over 
                  6 \times \mbox{Total Volume}} a^2
\non \\
             &=& {1 \over 6} \rho a^2
\non \\
             &=& {1\over 6} {\rho \over \Lambda^2} 
                    \left( {6\pi^2 \over 11} \b \right)^{102/121} 
                    \exp\left[- {6\pi^2 \over 11} \b \right]
\label{p}
\eea
where $a$ is the lattice spacing.  The upshot is that
$p$, which is the fraction of plaquettes pierced by P-vortices 
(equals the probability that any given plaquette is pierced by a P-vortex)
should scale like the string tension.  

\begin{figure}[h]
\centerline{\scalebox{1.2}{\includegraphics{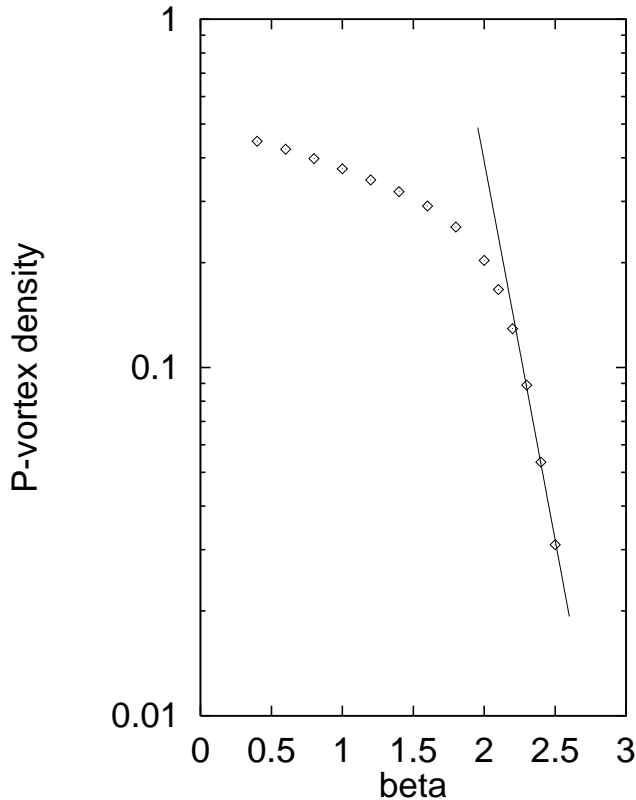}}}
\caption{Evidence for asymptotic scaling of the P-vortex density,
defined as the fraction $p$ of plaquettes pierced by P-vortices 
(one-sixth the average area occupied by P-vortices per unit lattice volume).
The solid line is the asymptotic freedom prediction of eq. \rf{p},
with constant $\protect \sqrt{\rho/(6\Lambda^2)} = 50$.}
\label{pvor}
\end{figure}

   Return now to the Creutz ratio plot in Fig.\ \ref{nchi}, and in particular 
the data for the center-projected $\chi(1,1)$, where
\beq
     \chi(1,1) = - \log W_{cp}(1,1)
\eeq
where the ``cp'' subscript indicates that this is the center-projected
Wilson loop.  It is easy to see that
\beq
         W_{cp}(1,1) = (1-p) + p\times (-1) = 1-2p
\eeq
so for small $p$ (large $\b$) we have
\beq
        \chi(1,1) \approx 2p
\label{x11}
\eeq
From the behavior of $\chi(1,1)$, which seems to (at least roughly)
parallel the straight line shown,  we see that $p$ does appear to scale
correctly. However, since \rf{x11} is approximate, it is better to plot
the precise value of the P-vortex density
\beq
       p = \oh (1 - W_{cp}(1,1))
\eeq
versus coupling $\b$, as shown in Fig.\ \ref{pvor}.  The straight line
is the asymptotic freedom expression (last line of eq.\ \rf{p}), with 
the choice $\sqrt{\rho/(6\Lambda^2)} = 50$.  The scaling of P-vortex densities,
at the larger $\b$ values, is rather compelling.  There seems little doubt
that P-vortices are locating physical, surface-like objects 
in the full Yang-Mills vacuum; objects which we have identified, in 
section 2.2 above, as center vortices.

\section{Cooled Vortices}        

   It has been argued persuasively by Teper \cite{Teper} that the
lattice cooling procedure can never, in a finite number of cooling steps, 
remove the asymptotic string tension extracted from sufficiently large
Wilson loops. However, as the number of cooling steps increases, the area-law
falloff sets in at increasingly large loop sizes; this means that for a 
lattice of any fixed volume, confinement is eventually lost. 
We would like to understand, in the vortex picture, how it is that the 
area law is lost for smaller loops while being preserved
for larger loops, after a finite number of cooling steps.   

\begin{figure}
\centerline{\scalebox{.6}{\includegraphics{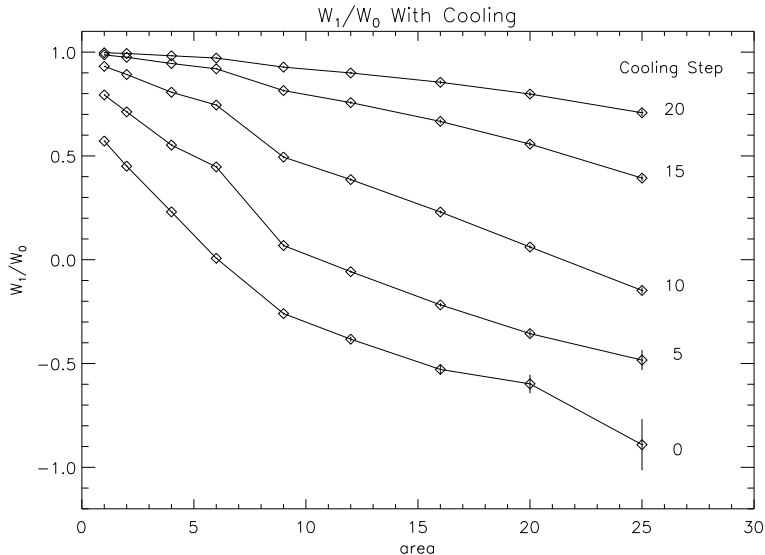}}}
\caption{Variation of the ratio $W_1/W_0$ with the number of cooling
steps.}
\label{w1w0}
\end{figure}

   To answer this question, we need to know what happens to vortices as
the lattice is cooled.  
As before, we use P-vortices found on the uncooled lattice to locate
the center vortices, and to count the number of times $n$ that vortices 
pierce a given lattice surface.  The lattice is then cooled, using the
constrained cooling procedure of Campostrini et al.\ \cite{Pisa}, and
we can study what has happened to the configurations identified on the
uncooled lattice.
The first quantity of interest is $W_1/W_0$, where the Wilson loops are 
evaluated on subensembles of the cooled, unprojected lattice, and the 
P-vortices are identified on the uncooled lattice.  The result, from 
0-20 cooling steps at $\beta=2.3$, is
shown in Fig.\ \ref{w1w0} (all data in this section was obtained on
a $16^4$ lattice).  A rough guide to the thickness of a
vortex is the loop size for which $W_1/W_0 \approx 0$. According to
Fig.\ \ref{w1w0}, this happens for $2\times 3$ loops at cooling step 0,
$3\times 4$ loops at cooling step 5, and $4\times 5$ loops at cooling
step 10.  The simplest interpretation is that the vortices become thicker
as the cooling proceeds, with $W_1/W_0$ reaching its asymptotic value
at ever larger distance scales.  

   The thickening of vortices with cooling explains how the area law
is lost for smaller loops, but retained asymptotically for sufficiently
large loops.  The asymptotic string tension is only obtained for loops
whose dimensions are significantly larger than the vortex thickness; there
is no area law falloff for loops whose size is very much \emph{smaller}
than the vortex thickness.\footnote{This point is discussed in much more 
detail in ref.\ \cite{castex}.}  Thus, as cooling begins, loops whose size is
comparable to the vortex thickness lose their area-law falloff
while the string tension of larger loops is unchanged.  As cooling
proceeds, vortex thickness increases, and the area law is lost for still
bigger loops.  However, after any number of cooling steps, there
will always be loops (on a sufficiently large lattice) whose extension
is large compared to the cooled vortex thickness, and whose asymptotic
string tension is untouched.

\begin{figure}
\centerline{\scalebox{.6}{\includegraphics{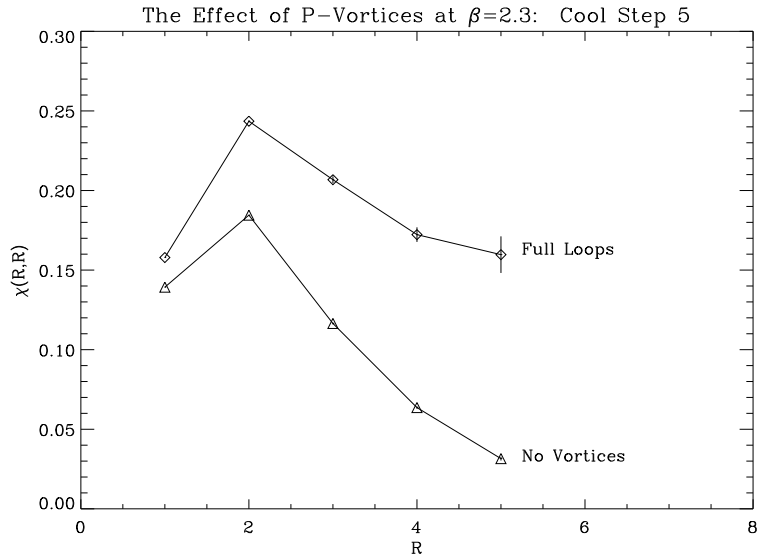}}}
\caption{Zero vortex Creutz ratio $\chi_0(R,R)$ and the full
Creutz ratio $\chi(R,R)$ vs.\ R, after 5 cooling steps.}
\label{chivsr5}
\end{figure}

\begin{figure}
\centerline{\scalebox{.6}{\includegraphics{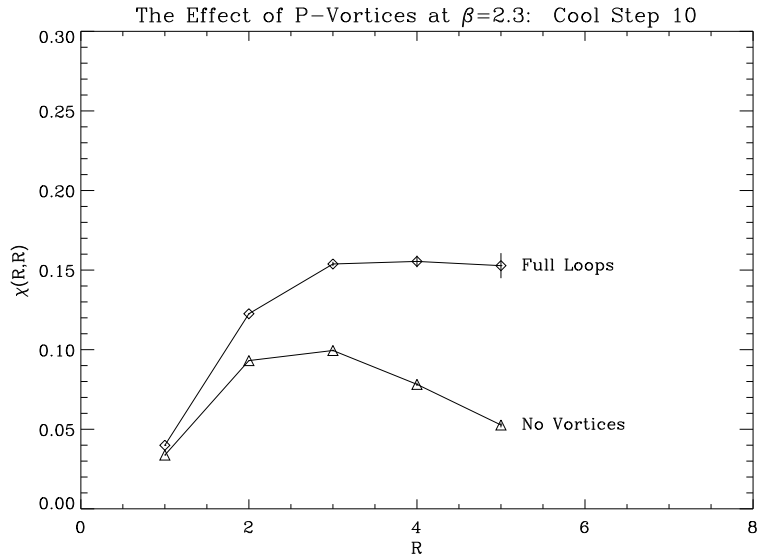}}}
\caption{Same as the previous figure, after 10 cooling steps.}
\label{chivsr10}
\end{figure}

   This picture is supported
by the data for $\chi_0(R,R)$ vs.\ $\chi(R,R)$ after 5 cooling steps
(Fig. \ref{chivsr5}), and after 10 cooling steps (Fig. \ref{chivsr10}),
again at $\b=2.3$.  These figures should be compared with 
Fig.\ \ref{nchi0} above, which shows the same quantities on an uncooled 
lattice.   We notice, particularly after 10 cooling steps, that the
Creutz ratios for small loops have been drastically reduced.  However, as
$R$ increases, the standard Creutz ratio $\chi(R,R)$ rises and seems to level
out near the usual value of the asymptotic string tension at $\beta=2.3$.
On the other hand, the zero-vortex Creutz ratio $\chi_0(R,R)$ is again
tending to zero for large loops.  At 0 cooling steps (Fig. \ref{nchi0}),
$\chi_0(R,R) \approx 0$ at $R=5$.  At 5 and 10 cooling steps, 
$\chi_0(5,5) > 0$, although the trend towards zero at increasing loop size is
clear.  A rough guess is that $\chi_0(R,R) \approx 0$ for $R$ such that
$W_1/W_0 \approx -1$.

   The message of figures \ref{w1w0}-\ref{chivsr10} is that
the vortices are still present on cooled lattices, and are still
essential to confinement.  However, the asymptotic values
$W_1/W_0 \ra -1$ and $\chi_0 \ra 0$ are obtained only at increasingly
large loop area, as the number of cooling steps increases.  This
behavior, as well as the loss of area law falloff for smaller loops with
cooling, seems to be nicely explained as being due to the ``thickening'' of 
the vortex core (which is the region of the center vortex which cannot be 
represented by a gauge transformation with discontinuity \rf{discont}).  

\begin{figure}
\centerline{\scalebox{.6}{\includegraphics{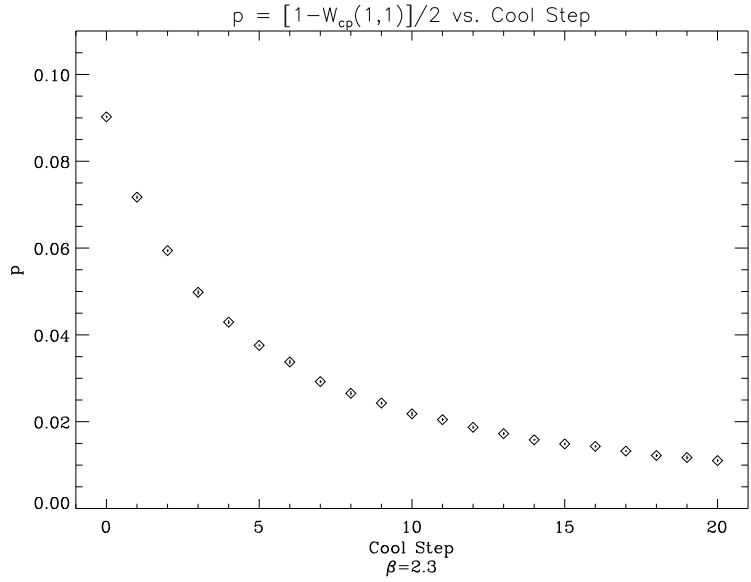}}}
\caption{The drop in P-vortex density $p$, identified by
center-projection in maximal center gauge, on cooled lattices.}
\label{pmeth1}
\end{figure}

\begin{figure}
\centerline{\scalebox{.6}{\includegraphics{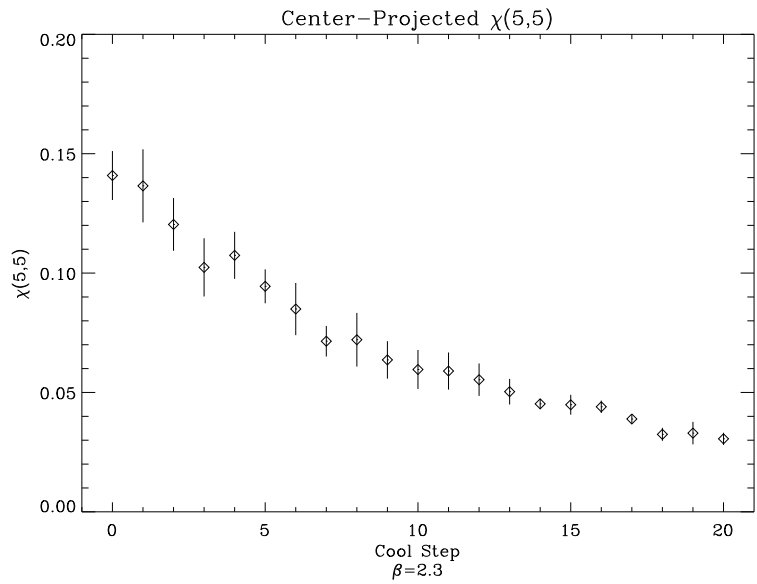}}}
\caption{The drop in the center-projected Creutz ratio $\chi_{cp}(5,5)$, 
corresponding to the drop in $p$, with cooling step.}
\label{cz55}
\end{figure}

   As just explained, our strategy is to locate the configurations
of interest on the uncooled lattice, and then study what happens to
these configurations as the lattice is cooled.  But it is also interesting
to ask whether our procedure for finding the center vortices, i.e. maximal
center gauge combined with center projection, also works on the
cooled lattices.  The answer is ``no.'' 
In Fig. \ref{pmeth1} we show how the P-vortex density $p$
of eq. \rf{p} falls drastically with cooling, if the P-vortices are
identified by gauge-fixing and center-projecting the cooled lattice.
A corresponding falloff is found in the Creutz ratios $\chi_{cp}(R,R)$
of center-projected loops, with increasing cooling step, as seen,
e.g., in Fig. \ref{cz55} for $R=5$.  There is no trace of the typical
``plateau'' in string tension over some finite number of cooling
steps, found in plotting the usual Creutz ratios.  We have also computed
$\chi_{cp}(R,R)$ on lattices that have been ``smoothed'' according to
the procedure in ref. \cite{DeGrand}, and find that the center-projected 
Creutz ratios are reduced by about a factor of three on the smoothed 
lattices, as compared to the original lattices.\footnote{We thank 
T. Kov\'{a}cs for kindly supplying us with 100 smoothed lattice 
configurations.}  Very similar phenomena have been reported
in maximal abelian gauge, in refs. \cite{HT,K}.

   We stress, however, that the center vortices themselves, on a large lattice,
are not removed by cooling.  That fact seems evident in Fig. 
\ref{w1w0}-\ref{chivsr10}.  What \emph{is} lost on the cooled lattice
is the efficacy of center-projection in finding all of the vortices.
This suggests that our procedure for locating center vortices is sensitive
not only to long-range fluctuations, but also to short-range features of some
kind that are associated with these vortices.  It would be interesting to
know what these short-range features are.\footnote{One clue is that the 
unprojected plaquette energy at the location of P-vortices, in the uncooled 
lattice, is significantly higher than the average plaquette energy.} This 
question is under investigation, and we hope to return to it in a future 
publication.

\section{Gribov Copies}

   The over-relaxation method of gauge-fixing, described in section 2,
is not guaranteed to find the absolute maximum of the quantity $R$
in eq. \rf{R}; in general it will only find a local maximum.  This is
the well known ``Gribov problem,'' which also afflicts the Coulomb,
Landau, and maximal abelian gauges.  It was in order to alleviate
the problem that we have made three random gauge copies of each configuration 
used for data taking, and gauge-fixed each to obtain three ``Gribov copies.''
We then used the Gribov copy with the largest value of $R$.  

   Since Gribov copies of a given lattice configuration are not 
identical, it is interesting to study by how much, on average, they differ.
In particular, to what extent are the positions of P-vortices correlated
from copy to copy?  If there is no correlation, then we can hardly rely
on P-vortices to locate physical objects (i.e. the thick center vortices).
On the other hand, some variation in the position of P-vortices, from
copy to copy, should be permissible.  Center vortices are rather
thick, extended objects, and the precise ``middle'' surface of such 
configurations, which P-vortices are supposed to locate, may be somewhat 
ill-defined. 

   To investigate quantitatively the correlation of vortices in different
Gribov copies, we do the following:  A Monte Carlo simulation is
run at a given $\b$, taking data every 100 sweeps, and
making 4 copies of each configuration chosen for data taking.  The
two ``best'' Gribov copies (in the sense of having the largest $R$ values)
are center-projected, and we denote the projected configurations
by $Z'_\m(x)$ and $Z''_\m(x)$, with corresponding center-projected
Wilson loops
\bea
            \W'(C)  &=& Z'Z'...Z'
\non \\
            \W''(C) &=& Z''Z''...Z''
\eea
Then we compute the expectation value of the loop product
\beq
          <\W'(C) \W''(C)>
\label{lprod}
\eeq
If the correlation of P-vortices in the two Gribov copies were
perfect, then
\beq
             \W'(C) = \W''(C) = \pm 1
\eeq
and therefore
\beq
            <\W'(C) \W''(C)> = 1  ~~~~~ \mbox{perfect correlation}
\eeq
At the other extreme, if there were no correlation at all between the
P-vortex positions in the two Gribov copies, then
\bea
  <\W'(C) \W''(C)> &=& <\W'(C)> <\W''(C)> 
\non \\
                   &=& \exp[-2\s \mbox{Area}(C)]
                             ~~~~~ \mbox{no correlation}
\eea
where $\s$ is the string tension of the center-projected loops (same
as the asymptotic tension of the unprojected loops).
            
   Small variations in P-vortex position among different Gribov copies 
are most likely to lead to a perimeter law falloff of the loop
product \rf{lprod}, at least in the limit of large loop area.  Consider, for 
example, the following simple model:  Take an $I\times J$ Wilson
planar loop and assign to each of the plaquettes in its plane, both
inside and outside the loop,
a value $+1$ with probability $(1-f)$, and $-1$ with probability $f$.
Each such configuration is supposed to represent a particular ``Gribov copy'' 
of center-projected plaquettes in the plane, and the value of the
Wilson loop is $\W'_{cp}(I,J)=(-1)^{n'}$, where $n'$ is the number of
negative plaquettes in the minimal area.  It is not hard to see,
since the plaquettes are assumed to be uncorrelated, that, averaging
over many configurations gives
\bea
       W'_{cp}(I,J) &=& <\W'(I,J)>
\non \\
                &=&  (1-2f)^{IJ} = \exp[-\s IJ]
\eea
where $\s = -\log(1-2f)$ (the assumption that nearby plaquettes are
completely uncorrelated is the main unrealistic feature of this model). 
From a given configuration of $\pm 1$ plaquettes, 
we construct a second ``Gribov copy'' by allowing negative plaquettes to
change their position by, at most, one lattice
spacing.  Then only changes in position of negative plaquettes bordering the 
loop perimeter can cause the loop product to differ from $+1$. Assign to 
a negative plaquette on the perimeter a probability $q$ to cross \emph{into}
the loop, if it were outside, or to cross outside the loop, if it were 
inside.\footnote{Plaquettes touching the corners of the loop should be treated
a little differently from the other plaquettes along the perimeter, but this 
is an inessential complication of the model, which we will ignore.}
The new value of the loop is $(-1)^{n''}$, where $n''$ is the
number of negative plaquettes inside the loop in the second ``Gribov copy.''
One then finds that in this model, defining $N \equiv 4(I+J)$,
\bea
      <\W'(I,J) \W''(I,J)> &=& \sum_{n=0}^N (1-2q)^n f^n (1-f)^{N-n}
                                {N! \over n! (N-n)!}
\non \\
           &=& (1-2qf)^{4(I+J)}
\eea
which is a  perimeter-law falloff.  

   The above argument should also go through if the P-vortex positions
vary by more than one lattice spacing among Gribov copies, so long as
the variation is small compared to the size of the loop.  If the variation
in P-vortex position is comparable to the thickness of the center vortex,
then our best chance to see perimeter-law falloff in the loop product
\rf{lprod}, for comparatively small-size loops, will be at smaller values
of coupling $\b$, where the vortex is relatively thin in lattice units.
We have therefore chosen to do our simulation at a value of $\b=2.1$
which is just past the strong-to-weak coupling crossover.

\begin{figure}
\centerline{\scalebox{.6}{\includegraphics{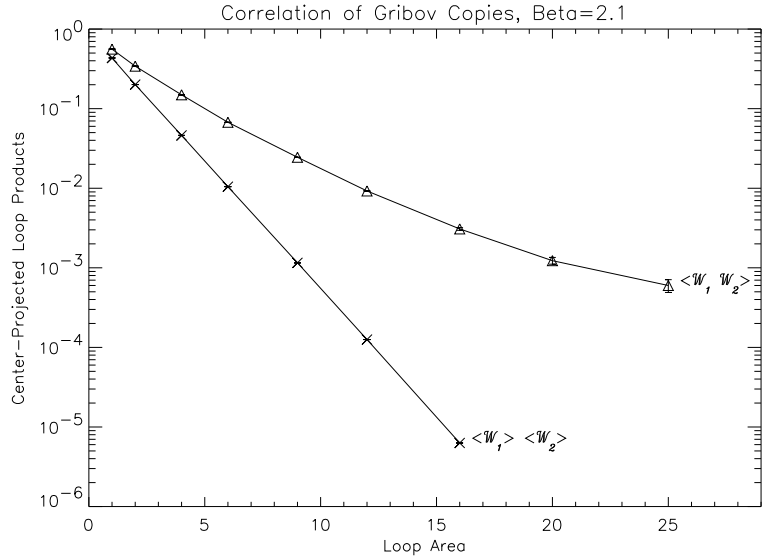}}}
\caption{Expectation value of products of center-projected Wilson
loops (triangles), evaluated in different Gribov copies, plotted 
vs.\ Loop Area.  Crosses indicate the value for no correlation.}
\label{zcopyA}
\end{figure}

\begin{figure}
\centerline{\scalebox{.6}{\includegraphics{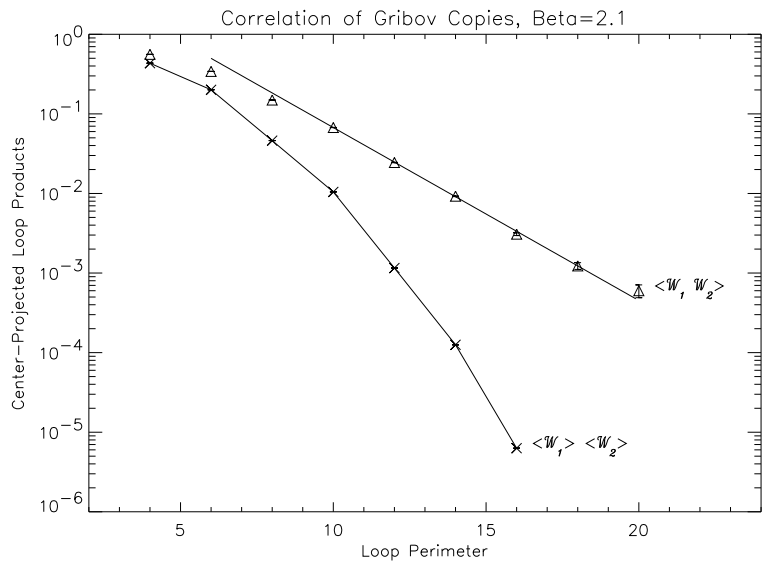}}}
\caption{Expectation value of products of center-projected Wilson
loops, evaluated in different Gribov copies, plotted vs.\ Loop Perimeter.}
\label{zcopyP}
\end{figure}

   Figures \ref{zcopyA} and \ref{zcopyP} show our Monte Carlo results
at $\b=2.1$ on a $14^4$ lattice, with data from 400 configurations
separated by 100 sweeps between configurations.  Triangles show the data for 
loop products $<\W'(C) \W''(C)>$, plotted vs.\ loop area, while the crosses 
are the values for no correlation, i.e. $<\W'(C)><\W''(C)>$.  The loop 
products are clearly far above the uncorrelated value and, from Fig.\
\ref{zcopyA}, do not seem consistent with an area-law falloff.  In Fig.\
\ref{zcopyP} the loop product is plotted vs.\ loop perimeter.  The straight
line is drawn, somewhat arbitrarily, through data points at perimeter 
$= 10,~18$.  It
appears that the falloff in the loop product with perimeter is quite
compatible with perimeter-law falloff, as predicted in our simple model.

   These results indicate that the variation in P-vortex position among
different Gribov copies is relatively small - perhaps on the order of
the vortex thickness, although we have not quantified this - and leads
asymptotically only to a perimeter-law falloff for the loop product 
$<\W'(C) \W''(C)>$, indicating a strong correlation among Gribov copies.

\section{First Results in SU(3)}
All results presented in the previous sections support the idea that
thick $Z_2$ vortices are {\em the} configurations dominating the
$SU(2)$ Yang--Mills vacuum. However, the vortex mechanism should not
be specific to the $SU(2)$ gauge group.  In nature quarks appear in three
colors, so a very urgent question is whether the observed phenomena
survive the transition from $SU(2)$ to $SU(3)$.

The maximal center gauge in $SU(3)$ gauge theory is defined as the gauge
which brings link variables $U$ as close as possible to elements of
its center $Z_3 = \lbrace e^{-2i\pi/3}I,\;I,\;e^{2i\pi/3}I\rbrace$. 
This can be achieved e.g.\ by maximizing the quantity
\begin{equation}\label{baryonlike}
R=\sum_{x,\mu}\mbox{Re}\left(\left[\mbox{Tr}\;U_\mu(x)\right]^3\right),
\end{equation}
or
\begin{equation}\label{mesonlike}
R'=\sum_{x,\mu}\mid\mbox{Tr}\;U_\mu(x)\mid^2.
\end{equation}
We will concentrate here on the first choice.
Center projection then amounts to replacing full link variable $U_\mu(x)$
by $Z_\mu(x)$, the closest center element. The residual unfixed local
gauge symmetry is that of $Z_3$.

Fixing to the maximal center gauge in $SU(3)$ gauge theory turns out to be 
much more difficult and computationally intensive than in the case of 
$SU(2)$. The reason is that we have not succeeded
in reducing the maximization to an underlying linear algebra problem 
as in $SU(2)$ (see Section 2.1). We thus resorted to the method of
simulated annealing \cite{Cerny,Kirkpatrick}, which was 
used for maximal abelian gauge fixing by Bali et al.\ \cite{Bali_simann}.
However, this method of maximal center gauge fixing converges to the 
maximum of $R$, Eq.\ (\ref{baryonlike}), very slowly, which has forced us 
thus far to restrict simulations to small lattice sizes and to strong coupling.
Tests of a more efficient maximization procedure are in progress.

\begin{figure}
\centerline{\scalebox{.65}{\includegraphics{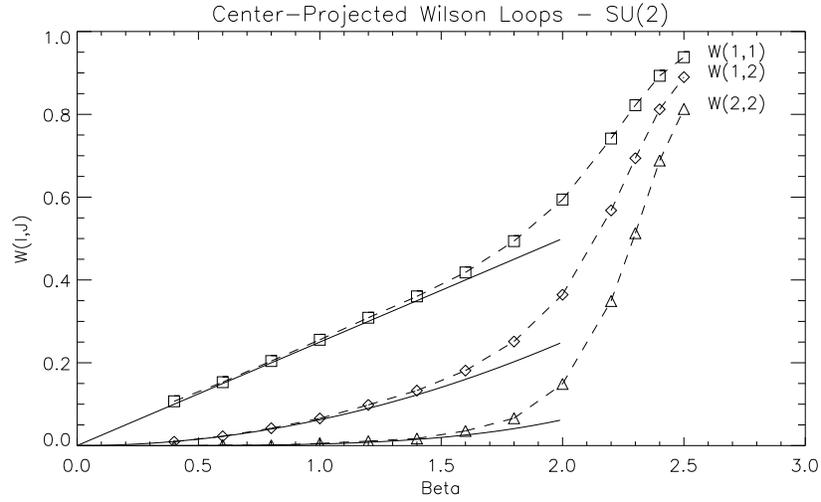}}}
\caption{Center-projected Wilson loops vs.\ the strong-coupling expansion
(solid lines) in $SU(2)$ lattice gauge theory}
\label{SU2strong}
\end{figure}

Before discussing the strong coupling results for $SU(3)$, let us first
show analogous data from $SU(2)$ gauge theory. In Figure \ref{SU2strong}
we plot values of center-projected Wilson loops $W(I,J)$ in 
maximal center gauge for 
$\beta\le2.5$. Broken lines connecting data points are just meant
to guide the eye. Solid lines represent result of the lowest-order
strong-coupling expansion (for unprojected loops)
\begin{equation}
W(I,J)=\left(\frac{\beta}{4}\right)^{IJ}.
\end{equation}
Monte Carlo data for projected loops agree with the lowest-order strong 
coupling expansion up to about $\beta=1.5$.

\begin{figure}
\begin{center}
\centerline{\scalebox{.65}{\includegraphics{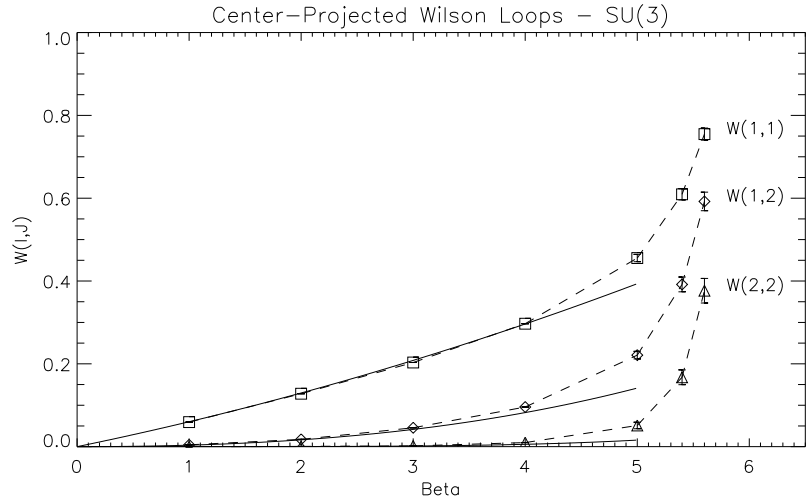}}}
\end{center}
\caption{Center-projected Wilson loops vs.\ the strong-coupling expansion
(solid lines) in $SU(3)$ lattice gauge theory}
\label{SU3strong}
\end{figure}

Our first results from $SU(3)$ lattice gauge theory simulations come
from an $8^4$ lattice, for $\beta$ values 1.0, 2.0, 3.0, 4.0, 5.0, 5.4 
and 5.6. Figure \ref{SU3strong} shows center-projected Wilson loops 
together with the standard strong-coupling expansion to leading and 
next-to-leading order:
\begin{equation}
W(I,J)=\left(\frac{\beta}{18}\right)^{IJ}\left(1+\frac{IJ}{12}\beta 
+O(\beta^2)\right).
\end{equation}
The data agree with lowest-order strong-coupling expansion up to
$\beta\simeq 2$; when next-to-leading term is taken into account, 
the agreement extends up to $\beta=4$.

Qualitatively, the situation at strong coupling looks much the same
in $SU(2)$ and $SU(3)$: in both cases full Wilson loops are well reproduced
by those constructed from center elements alone in maximal
center gauge. Thus, center dominance is seen also in $SU(3)$ gauge theory
at strong coupling.

An immediate task for the near future is to repeat our investigation of 
center dominance and the role of vortices in $SU(3)$ lattice gauge theory
for couplings in the scaling region.  An absolutely crucial check of the
validity of the vortex mechanism is that the evidence for vortices found in 
the $SU(2)$ lattice theory is also found for the $SU(3)$ gauge group.

\section{Summary}

   It may be worth summarizing the results reported here:

\begin{itemize}

\item{ {\bf P-Vortices locate center vortices.~~} Vortex excitations
in the center-projected configurations, in direct maximal center gauge,
locate center vortices in the full, unprojected lattice.  The
evidence for this comes from the fact that \newline
$W_n(C)/W_0(C) \ra (-1)^n$, and $W_{odd}(C)\ra -W_{evn}(C)$ in the limit
of large loop area.}

\item{ {\bf No vortices $\Rightarrow$ no confinement.~~}  When Wilson
loops in $SU(2)$ gauge theory  are evaluated in subensembles of 
configurations with no vortices (or only an even number of vortices) 
piercing the loop, the string tension disappears.} 

\item{ {\bf Vortices, by themselves, account for the full string tension.~~}
The string tension of the vortex contribution to Wilson loops is found
to match, quite accurately, the asymptotic string tension extracted from
the full Wilson loops.}

\item{ {\bf Vortex density scales.~~} The variation of P-vortex density
with coupling $\b$ goes as expected for a physical quantity with dimensions
of inverse area.  This is additional evidence that P-vortices locate 
physical, surface-like objects (center vortices) in the Yang-Mills vacuum
(see also ref.\ \cite{LR}).}

\item{ {\bf Center vortices thicken as the lattice cools.~~}  This enables
us to explain how the area law falloff is lost, after a finite number of
cooling steps, for smaller loops, while the string tension remains 
unchanged for sufficiently large loops.}

\item{ {\bf P-vortex locations are correlated among Gribov 
copies. ~~} There appears to be only modest sensitivity in P-vortex location
to the choice of Gribov copy.}

\item{ {\bf SU(3).~~} There is preliminary evidence, on small lattices
and strong couplings, of center dominance also in $SU(3)$ lattice gauge
theory.}

\end{itemize}

It is also worth mentioning some other results reported in refs. 
\cite{Zako,castex}:

\begin{itemize}

\item{ {\bf Monopole loops lie on P-vortices \cite{Zako}.~~} 
Monopoles, identified in the maximal abelian gauge, lie along center vortices,
found in the indirect maximal center gauge, in a monopole-antimonopole 
chain.  The non-abelian field strength of monopole cubes, above the lattice 
average, is directed almost entirely along the associated center vortices. 
Monopoles appear to be rather undistinguished regions of vortices, and may 
simply be artifacts of the abelian projection, as explained in ref.\ 
\cite{Zako}.} 
     
\item{ {\bf Center vortices are compatible with Casimir 
scaling \cite{castex}.~~}  The ``Casimir scaling'' of the string tension of
higher representation Wilson loops, at intermediate distance scales,
has long been considered incompatible with the center vortex theory. Very 
recently, however, it has been argued that Casimir scaling is explained in 
terms of center vortices, if we take into account the fact that center 
vortices, unlike P-vortices, have a thickness which may be much greater than 
one lattice spacing.\footnote{Related work on the adjoint potential,
in the context of a particular model of center vortices, may be found in
ref.\ \cite{Corn2}; some speculations about hedgehog solutions, in the same
framework, are found in ref.\ \cite{Corn3}.  The approach taken in ref.\ 
\cite{Corn2} has some similarities to ours in ref.\ \cite{castex}, but also
differs in a number of important respects. In 
particular there is no apparent Casimir scaling found in the former approach, 
and there also seems to be an explicit conflict with the large-N factorization 
property at $N_{colors}\ra \infty$.} }

\end{itemize}

   These results support the view that center vortices are responsible 
for quark confinement.

\vspace{33pt}

\ni {\Large \bf Acknowledgements}

\bigskip

   This work was supported in part by PPARC under Grant GR/L56329 (L.DD.),
Fonds zur F\"orderung der Wissenschaftlichen Forschung P11387-PHY (M.F.), 
the U.S. Department of Energy under Grant No. DE-FG03-92ER40711
and Carlsbergfondet (J. Gr.), the ``Action Austria-Slovak Republic: 
Cooperation in Science and Education'' (Project No. 18s41) and the Slovak 
Grant Agency for Science, Grant No. 2/4111/97 (\v{S}. O.).  

   J.Gr. and J.Gi. are grateful for the generous assistance of the 
High Energy Theory Group at Lawrence Berkeley National Laboratory, whose 
computer facilities were used extensively in this project.


\begin{thebibliography}{xx}
\bibitem{PRD97}L. Del Debbio, M. Faber, J. Greensite, and 
{\v S}. Olejn\'{\i}k, Phys. Rev. D55 (1997) 2298, hep-lat/9610005.
\bibitem{Zako}L. Del Debbio, M. Faber, J. Greensite, and 
{\v S}. Olejn\'{\i}k, proceedings of the Zakopane meeting
{\sl New Developments in Quantum Field Theory}, hep-lat/9708023; \\
proceedings of the Buckow meeting {\sl 31st Int. Symposium on the
Theory of Elementary Particles}, in preparation.
\bibitem{lat97}L. Del Debbio, M. Faber, J. Greensite, and 
{\v S}. Olejn\'{\i}k, proceedings of LATTICE 97, hep-lat/9709032.
\bibitem{tHooft}G. 't Hooft, Nucl. Phys. B138 (1978) 1. 
\bibitem{Mack}G. Mack, in {\sl Recent Developments in Gauge Theories},
edited by G. 't Hooft et al. (Plenum, New York, 1980).
\bibitem{Cop}H. B. Nielsen and P. Olesen, Nucl. Phys. B160 (1979) 380; \\
J. Ambj{\o}rn and P. Olesen, Nucl. Phys. B170 (1980) 60; 265.
\bibitem{Corn}J. M. Cornwall, Nucl. Phys. B157 (1979) 392.
\bibitem{Feyn}R. P. Feynman, Nucl. Phys. B188 (1981) 479.
\bibitem{castex}M. Faber, J. Greensite, and {\v S}. Olejn\'{\i}k,
hep-lat/9710039, to appear in Phys.\ Rev.\ D.  
\bibitem{TK}T. Kov\'{a}cs and E. Tomboulis, hep-lat/9711009; 
hep-lat/9709042. 
\bibitem{Suz} A. Kronfeld, M. Laursen, G. Schierholz, and U.-J. Wiese, \\
Phys. Lett. B198 (1987) 516 ;\\
T. Suzuki and I. Yotsuyanagi, Phys. Rev. D42 (1990) 4257.
\bibitem{lat96}L. Del Debbio, M. Faber, J. Greensite, and 
{\v S}. Olejn\'{\i}k, Nucl. Phys. Proc. Suppl. 53 (1997) 141, 
hep-lat/9607053.
\bibitem{MO} J. E. Mandula and M. Ogilvie, Phys. Lett. B248 (1990) 156.
\bibitem{Bali}G. Bali, C. Schlichter, and K. Schilling, Phys. Rev. D51
(1995) 5165.
\bibitem{LR}K. Langfeld, H. Reinhardt, and O. Tennert,
hep-lat/9710068.
\bibitem{Teper} M. Teper, Nucl. Phys. B411 (1994) 855.
\bibitem{Pisa} M. Campostrini, A. Di Giacomo, M. Maggiore, H. Panagopoulos,
and E. Vicari, Phys. Lett. B225 (1989) 403.
\bibitem{DeGrand} T. DeGrand, A. Hasenfratz, and T. Kov\'{a}cs,
Nucl. Phys. B505 (1997) 417, hep-lat/9705009.
\bibitem{HT} A. Hart and M. Teper, hep-lat/9709009.
\bibitem{K} T. G. Kov\'{a}cs and Z. Schram, Phys. Rev. D56 (1997)
6824, hep-lat/9706012.
\bibitem{Cerny}
V.\ \v{C}ern\'y, Comenius Univ.\ preprint (1982),
J.\ Opt.\ The.\ Appl.\ 45 (1985) 41.
\bibitem{Kirkpatrick}
S.\ Kirkpatrick, C.\ D.\ Gelatt, Jr., M.\ P.\ Vecchi, Science 220 (1983) 671.
\bibitem{Bali_simann}
G.\ S.\ Bali, V.\ Bornyakov, M.\ M\"uller-Preussker, and F.\ Pahl,
Nucl.\ Phys.\ Proc.\ Suppl. 42 (1995) 852.
\bibitem{Corn2}J. M. Cornwall, hep-th/9712248; \\
and in Proceedings of the {\sl Workshop on Non-Perturbative Quantum 
Chromodynamics}, edited by K. A. Milton and M. A. Samuel (Birkhauser, 
Boston, 1983).
\bibitem{Corn3}J. M. Cornwall and G. Tiktopoulos, Phys. Lett. B181 (1986) 353.
\end{thebibliography}
\end{document}